\documentclass[prc,fleqn,twoside,floatfix]{revtex4}
\usepackage{latexsym,exscale}
\usepackage{epsfig}
\usepackage{amsmath,amssymb,amsfonts}
\begin{document}
\title{Mean field and pairing properties in the crust of neutron stars}
\author{F. Montani}
\affiliation{Institut d' Astronomie et d' Astrophysique, \\
Universite Libre de Bruxelles, Campus de la Plaine CP-226, \\
Boulevard du Triomphe, B-1000 Brussels, Belgium}
\author{C. May and H. M\"uther}
\affiliation{Institut f\"ur
Theoretische Physik, \\ Universit\"at T\"ubingen, D-72076 T\"ubingen, Germany}
 
\begin{abstract}
Properties of the matter in the inner crust of a neutron star are investigated in
a Hartree-Fock plus BCS approximation employing schematic effective forces of
the type of the Skyrme forces. Special attention is paid to differences 
between a homogenous and inhomogeneous description of the matter distribution.
For that purpose self-consistent Hartree Fock calculations are performed in a
spherical Wigner-Seitz cell. The results are compared to predictions of
corresponding Thomas Fermi calculations. The influence of the shell structure on
the formation of pairing correlations in inhomogeneous matter are discussed.
\end{abstract}
%\pacs{21.60.-n,21.60.Jz,26.60.+c} 
\maketitle

\section{Introduction}
The determination of the equation of state (EoS) for nucleon matter is an
important ingredient for various investigations of astrophysical objects. A lot
of attention has been paid to the EoS of baryonic matter at densities above the
saturation density of nuclear matter ($\rho_0 \approx 0.16$ fm$^{-3}$). These
densities should be relevant to describe the interior of neutron stars and are
of particular interest as they give rise for speculations about exotic phases of
baryonic matter including the existence of kaon condensates or quark
matter\cite{prak1,heisel,grein1}. 

However, it is not only this regime of very high densities also the crust of
neutron stars should be a very intriguing phase of baryonic matter. At those
lower densities and low temperatures the free energy should be reduced by
allowing for a phase of inhomogeneous baryonic matter.  This nonuniform matter
should consist out of a lattice of quasi-nuclei embedded in a gas of electrons
and possibly also a sea of neutrons. Thomas Fermi calculations, which are based
on non-relativistic or relativistic mean field calculations for the homogenous
system, exhibit these features\cite{oyama,oyama2,shenn,pol1,watana}. 

The presence of neutron superfluidity in the crust of neutron stars seems to be
well established. This is based on investigations solving the BCS gap equation
in homogenous nuclear matter\cite{baldo,kodel,elgar,kuck}. The precise knowledge
of the pairing gap is very important for the determination of the cooling rate
of neutron stars. The superfluidity of the material of the crust should have a
significant influence on the rotation frequencies of the star. The formation of
glitches may be related to the vortex pinning of the superfluid phase in the
inhomogeneous crust.

The aim of the investigations presented in this paper is to explore basic
properties of this inhomogeneous phase of baryonic matter on a level beyond the
Thomas Fermi approximation. What are the consequences of shell effects in the
quasi-nuclei in the systems for the bulk-properties of this material? How do
they influence the proton fraction in the $\beta$ equilibrium? What are the
consequences of these quasi-nuclei on the pairing properties of the system?

To answer this kind of questions we perform nuclear structure calculations in a
Wigner-Seitz cell of spherical shape. The assumption of such a spherical cell is
not designed to allow for inhomogeneities of the matter which is different from
the formation of quasi-nuclei or bubbles in the homogenous matter. Therefore we
do not consider a formation of rods or slabs of increased density\cite{watana}.
Also at this stage we do not aim at a complete survey of this matter covering a
large region of various densities. Instead we wish to explore some features of
nuclear structure calculations beyond the Thomas Fermi approach for a few
selected examples.

For that purpose we perform Hartree-Fock and Hartree-Fock plus BCS calculations
in an appropriate basis of the spherical Wigner-Seitz cell. For the NN
interaction we use rather simple effective forces of the Skyrme
type\cite{skyr,brink}. Special attention is paid to the pairing gaps resulting
from calculations of the homogenous matter and the non-homogeneous system. 

After this introduction we discuss in section 2 the techniques and results of
Hartree-Fock calculations for the $\beta$-stable matter within a spherical
Wigner-Seitz cell. The determination of pairing properties is presented in
section 3, which is followed by a short summary and conclusions in section 4.

\section{Hartree-Fock calculations in a spherical box}
The single-particle wave functions for the nucleon considered in our 
calculations are expanded
in a complete basis of orthonormal states defined within a spherical box of
radius $R$. Such a orthonormal set of basis functions, which are regular at 
the origin in the center of the box is given by 
\begin{equation}
\Phi_{iljm}(\vec r) = \langle \vec r \vert iljm\rangle = R_{il}(r)
{\cal Y}_{ljm} (\vartheta,\varphi)\,.\label{eq:basis}
\end{equation}
In this equation ${\cal Y}_{ljm}$ represent the spherical harmonics including
the spin degrees of freedom by coupling the orbital angular momentum $l$ with
the spin to a single-particle angular momentum $j$. The radial wave functions
$R_{il}$ are given by the spherical Bessel functions, $R_{il}(r)\sim j_l(k_ir)$ 
for the discrete momenta $k_i$, which fulfill
\begin{equation}
R_{il}(R)= N_{il}j_l(k_iR) = 0\,.\label{eq:basis1}
\end{equation}
The normalization constant
\begin{equation}
N_{il}=\left\{\begin{array}{ll}\frac{\sqrt{2}}{\sqrt{R^3}j_{l-1}(k_iR)} &
\mbox{for}\,l>0\,,\\
\frac{i\pi\sqrt{2}}{\sqrt{R^3}}&\mbox{for}\,l=0\,,\end{array}\right.
\end{equation} 
ensures that the basis functions are orthogonal and normalized within the box,
\begin{equation}
\int_0^R d^3r \Phi^*_{iljm}(\vec r)\Phi_{i'l'j'm'}(\vec r) =
\delta_{ii'}\delta_{ll'}\delta_{jj'}\delta_{mm'}\,.
\end{equation}
Adopting this basis of eigenstates for the kinetic energy one can try to
describe homogenous nuclear matter with a constant density $\rho_\pi$ and
$\rho_\nu$ for protons and neutrons, respectively, by filling all basis states 
with momenta $k_i$ below the corresponding Fermi momentum $k_{F\pi}$ and
$k_{F\nu}$. The local density is then given by
\begin{equation}
\rho(r) = \sum_{ilj} \Theta(k_F-k_i) (2j+1)R_{il}^2(r)\,,\label{eq:locdens}
\end{equation}  
where $\Theta$ stands for the Heaviside function. Results for this local density
considering a spherical box with radius $R$=25 fm and trying to describe a
system with a density of 0.06 fm$^{-3}$ is displayed in Fig.~\ref{plot25}. From
this figure one can see that the local density is in reasonable agreement with
the mean value except at the border of the cavity at $r=R$. Since all wave
functions by construction are bound to disappear at this borderline, the same is
also true for the density.

In order to cure this deficiency one could try an alternative basis for the
radial functions by the boundary condition that the radial derivative vanishes
at the surface of the box 
\begin{equation}
\widetilde R_{il}(r) = \widetilde N_{il} j_l(\widetilde{k_i} r) \quad\mbox{with}
\quad \frac{\partial \widetilde{R_{il}}}{\partial r}(R)=0\,.\label{eq:basis2}
\end{equation}
So all these basis function exhibit an extremum at the surface of the 
Wigner-Seitz cell which leads to a maximum of the local density at $r=R$ if
these functions are employed in (\ref{eq:locdens}) as on can see from the curve
labeled ``Basis 2'' in Fig.~\ref{plot25}.  

Bonche and Vautherin\cite{bonche} suggested to use a mixed basis by employing
e.g basis states with the boundary condition (\ref{eq:basis1}) for states with
even $l$ and those with the boundary condition (\ref{eq:basis2}) for states
with odd orbital angular momentum $l$. This recipe cures the deviation of the
local density from the mean value at the surface (see also Fig.~\ref{plot25}).
Nevertheless the representation of the homogenous phase of nuclear matter
within a spherical box of finite size leads to slight fluctuations in the local
density. Furthermore one must be aware of shell effects, which are due to the
finite size of the spherical box considered.  

In order to explore the influence of these shell effects on the evaluation of
expectation values, we have considered spherical boxes of various radii and
calculated the binding energy per nucleon for homogenous nuclear matter using
the representation of the plane wave states discussed above. We have considered
values for the radius $R$ ranging from 15 fm to 25 fm. Most of the results
turned out to be rather insensitive to the choice of the basis
((\ref{eq:basis1}), (\ref{eq:basis2}) or mixture). Therefore, if not stated
differently,  we will present results for the basis (\ref{eq:basis1}) only.

As a first example we consider the homogenous system of neutrons, protons and
electrons in $\beta$ equilibrium. This means that for any value of the baryon
density considered
\begin{equation}
\rho = \rho_{\pi} + \rho_{\nu}\,,
\end{equation}
we determine the proton abundance
\begin{equation}
Y_{\pi} = \frac{\rho_{\pi}}{\rho}\,,
\end{equation}
by the requirement that the Fermi energy for the neutrons is identical to the 
Fermi energy of the protons plus the Fermi energy of the electrons with the
density of electrons being identical to the density of protons.

\begin{figure}
\begin{center}
\epsfig{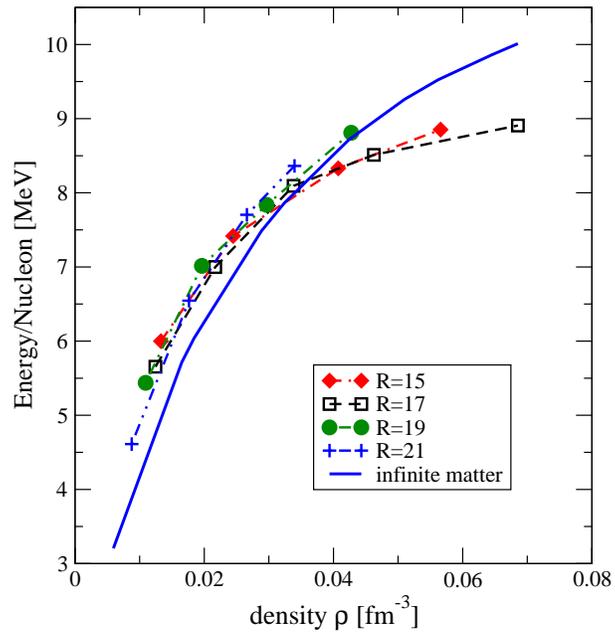}
\end{center}
\caption{\label{fig:figebet}Energy per nucleon for homogenous matter in $\beta$
equilibrium. Results for infinite matter are compared to those obtained in
spherical boxes of different radii. The Skyrme I has been used for the NN
interaction.}
\end{figure}

\begin{figure}
\begin{center}
\epsfig{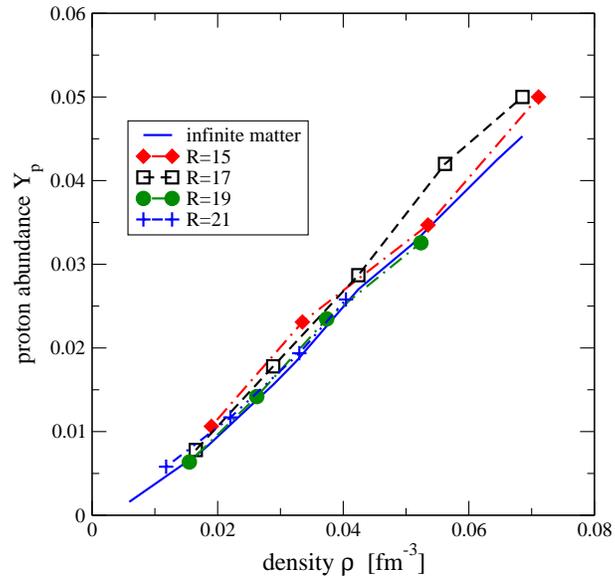}
\end{center}
\caption{\label{fig:figybet}Proton abundance $Y_p$ for homogenous matter in 
$\beta$ equilibrium. Further comments see Fig.~\protect{\ref{fig:figebet}}.}
\end{figure}

Results for the energy per nucleon and the proton abundance are displayed in
figures \ref{fig:figebet} and \ref{fig:figybet}, respectively. For the range of
densities considered the results obtained in the spherical boxes of different
size agree rather well with the corresponding observables calculated in the
infinite system. A significant discrepancy is only observed in the calculation
of the total energy at higher  densities considering boxes with small radii. In
these cases the surface effects displayed in Fig.\ref{plot25} as well as the
shell effects lead to energies which are to small as compared to the result for
the infinite system.   

\begin{figure}
\begin{center}
\epsfig{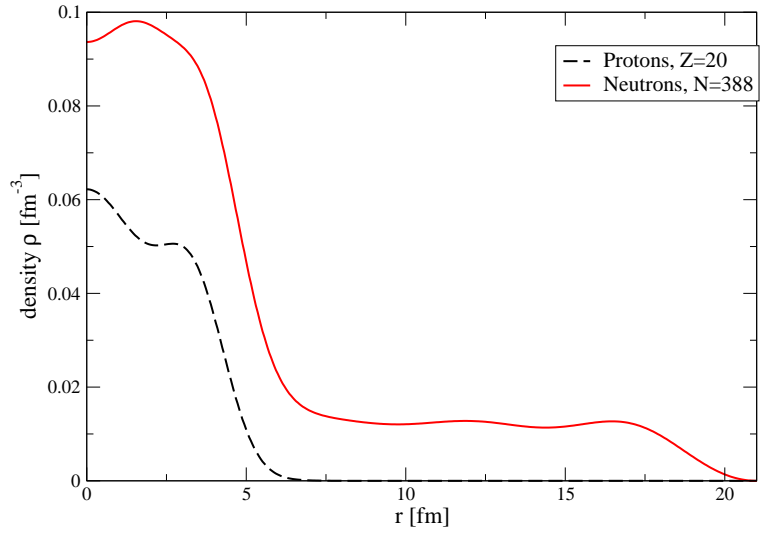}
\end{center}
\caption{\label{fig:rho1}Density distributions derived from Hartree
Fock calculations for nuclear matter in $\beta$ equilibrium.}
\end{figure}

After this test of the box basis to describe the homogenous system, we now turn
to the Hartree-Fock (HF) description within such a spherical box. The HF
single-particle wave functions are expanded in the basis of (\ref{eq:basis})
by
\begin{equation}
\Psi_{\alpha ljm}(\vec r) = \sum_{i=1}^Nc_{\alpha ilj}
\Phi_{iljm}(\vec r)\,.
\label{eq:hfbasis}
\end{equation}
The number of basis states $N$ is chosen to guarantee that the results are not
affected by this limitation. Results of self-consistent HF calculations are
displayed in Fig.~\ref{fig:rho1} showing the density of protons and neutrons as
a function of the radial distance from the center of a box with radius $R$=21
fm. the particle numbers for protons ($Z$=20 in this example) and neutrons
($N$=388) have been determined to fulfill the condition of $\beta$ stability,
employing the Fermi energies evaluated in the HF calculation. 

The density distribution for the protons is different from zero only near the
center of the Wigner Seitz cell. In this region also the density of the neutrons
is considerably larger than in the rest of the spherical box. Therefore we can
interpret this configuration as a quasi-nucleus embedded in a neutron see.
The single-particle energy spectrum for the protons exhibit a clear shell 
structure up to energies around the Fermi energy. The density of states around
the corresponding Fermi energy is larger for the neutrons. In fact it is close
to the one of the homogeneous matter described in the spherical box of this size.

For the example of $Z$=20 protons and $N$=388 neutrons in a Wigner Seitz cell
with radius $R$=21 fm considered in Fig.~\ref{fig:rho1} we obtain a global
density of $\rho$ = 0.0105 fm$^{-3}$ and a abundance for the protons of $Y_p$ =
0.049. The energy per nucleon for this self-consistent solution of the HF
equations is about 2 MeV per nucleon below the energy which was obtained for the
homogenous distribution of matter in $\beta$ equilibrium.

\begin{figure}
\begin{center}
\epsfig{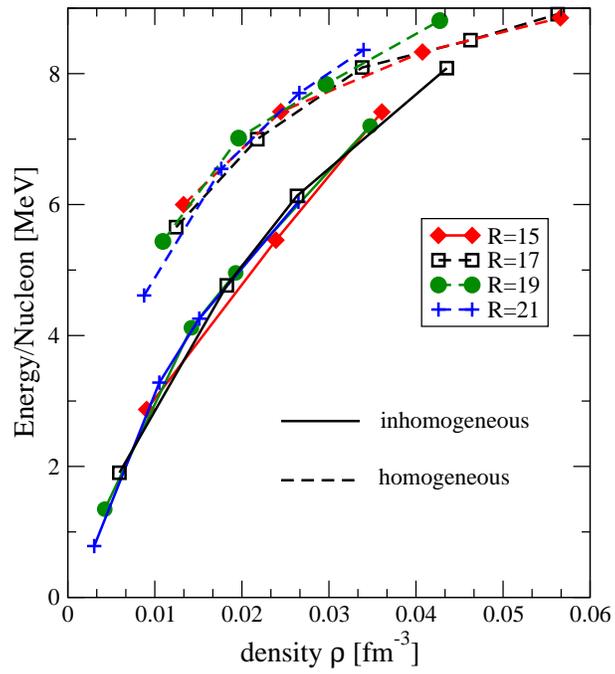}
\end{center}
\caption{\label{fig:enerinh}Energy per nucleon of $\beta$-stable nuclear matter
in the homogenous phase (symbols connected by dashed line) and inhomogeneous
phase (symbols connected by solid line).}
\end{figure}

Fig.~\ref{fig:enerinh} exhibits this gain in energy due to the formation of
quasi-nuclei in $\beta$-stable nuclear matter. For small densities around $\rho$
= 0.01 fm$^{-3}$ the HF solution with localized quasi-nuclei yields an energy
which is about 3 MeV per nucleon below the energy of the corresponding
homogenous matter. This gain in energy decreases with increasing density to
around 1 MeV per nucleon at  $\rho$ = 0.04 fm$^{-3}$. The results are rather
independent on the size of the Wigner Seitz cell under consideration.

\begin{figure}
\begin{center}
\epsfig{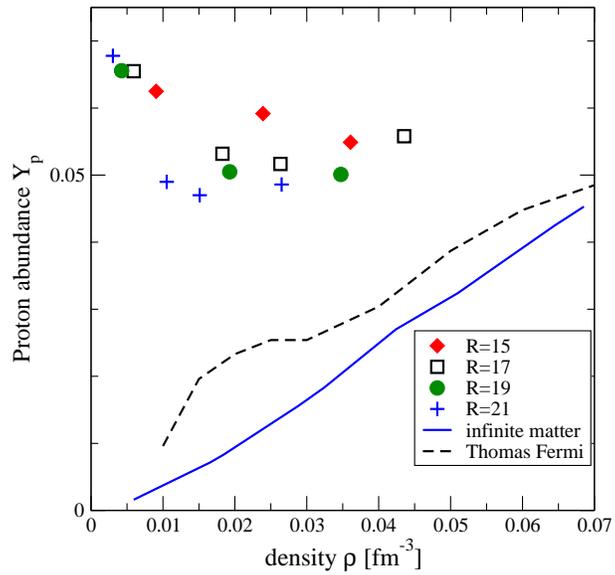}
\end{center}
\caption{\label{fig:ypinh}Proton abundances of $\beta$-stable nuclear matter
in the homogenous case of homogenous infinite matter (solid line), inhomogeneous
matter determined in the Thomas Fermi approximation and in Hartree-Fock
calculations employing Wigner Seitz cells of different radii.}
\end{figure}

The Hartree-Fock calculations allowing for an inhomogeneous distribution of
matter, however, also yield different results for the proton abundances as
compared to the results obtained for homogenous nuclear matter in $\beta$
equilibrium (see Fig.~\ref{fig:ypinh}). Thomas Fermi calculation, which allow for
an inhomogeneous distribution of matter, show proton abundances which are
slightly above those derived from the calculation of homogenous matter. This
indicates that the fluctuations in the density, which are taken into account in
the Thomas Fermi calculations, tend to enhance the proton abundances, as regions
with high densities contain a larger fraction of protons. However, the main
effect in the enhancement of the proton abundances observed in the Hartree Fock
approach is not described by the Thomas Fermi model. It originates from the
pronounced shell structure of the proton single-particle energies. The
Hartree-Fock calculation favors distributions of matter with a quasi-nucleus 
showing a closed shell for the protons.

\section{Pairing in the crust of neutron stars}
After we have noticed the effects of shell structure on the decomposition of the
nuclear material, we would like to address in this section the question how this
shell structure can affect the pairing properties of the nuclear material.
In particular we consider neutron-neutron pairing for neutron pairs with total
momentum equal to zero in $^1S_0$ partial wave for the relative motion. Using
the standard BCS approach for homogenous matter the pairing gap $\Delta (k)$ for
a pair of neutrons with relative momentum $k =\vert \vec k\vert$ is obtained by
solving the gap equation\cite{baldo,elgar,kuck}
 
\begin{equation}
\Delta(k)= -\frac{2}{\pi} \int_0^\infty dk'\,k'^2\, V(k,k') \frac{\Delta(k')}
{2\sqrt{\left(\varepsilon_k' - \varepsilon_F\right)^2 + \Delta(k')^2}}\,.
\label{eq:gap}
\end{equation}
 
Here $V(k,k')$ denotes the matrix elements of the NN interaction in the $^1S_0$ 
partial wave, $\varepsilon_k$ the single-particle energy for a nucleon with
momentum $k$ and $\varepsilon_F$ the Fermi energy. Instead of using the matrix
elements for a realistic NN interaction which is fitted to the NN scattering
data, we have chosen to employ an effective interaction which is of zero range,
density dependent and has been proposed by Bertsch and Esbensen
\cite{bert1,bert2}
\begin{equation}
V(r_1,r_2)=v_0 \left[1 -  \eta \left(\frac{\rho((r_1)}{\rho_0}\right)
^\alpha \right] \delta(r_1-r_2)\,,\label{eq:vdelta}
\end{equation}
where $v_0$, $\eta$ and $\alpha$ are adjustable parameters and $\rho_0$ is the
saturation density of nuclear matter. For such a
zero-range interaction a cut-off parameter $\varepsilon_C$ must be introduced in
the gap equation to truncate the integral to states with energy $\varepsilon_k$
less than $\varepsilon_C$. These four parameters can be constrained to reproduce
the $NN$ scattering length and Garrido et al.\cite{garrido} determined various 
sets of parameters which reproduce pairing gaps for nuclear and neutron matter
calculated with realistic NN interactions. We used the parameters $v_0=481$ 
MeV fm$^3$, $\eta=0.45$, $\alpha=0.47$ and $\varepsilon_C$ = 60 MeV and verified
that this set of parameters leeds to pairing gaps for homogenous neutron matter,
which are in fair agreement with those evaluated for the CD Bonn
interaction\cite{cdbonn}. 

We now turn to the solution of the HF+BCS equations\cite{schuck}, which in the 
spherical Wigner-Seitz cell have the form
\begin{eqnarray}
\left(\varepsilon_{nlj}-\varepsilon_{F}\right)  u_{nlj} & + \Delta_{nlj} 
v_{nlj} & = E_{nlj} u_{nlj}\nonumber \\
-\left(\varepsilon_{nlj}-\epsilon_{F}\right)  v_{nlj} & + \Delta_{nlj} 
u_{nlj} & = E_{nlj} v_{nlj}\,,\label{eq:bcsm}
\end{eqnarray}
where
$$
E_{nlj} = \sqrt{(\varepsilon_{nlj}-\varepsilon_F)^2 + \Delta_{nlj}^2}\,,
$$
is the energy of the quasiparticle state with quantum numbers $n,l,j$ and 
$\varepsilon_{nlj}$ the corresponding single-particle energy determined in the
HF equations. From the coefficients $u_{nlj}$, $v_{nlj}$ and the corresponding
single-particle wave functions $\Psi_{nljm}$ (see eq.(\ref{eq:hfbasis})) one can
calculate the anomalous density
\begin{equation}
\chi(r) = \sum_{nlj} (2j+1) \frac{u_{nlj}v_{nlj}}{2} \Psi_{nlj}^2(r)
\,,\label{eq:chir}
\end{equation} 
and the expression for the normal density is changed into
\begin{equation}
\rho(r) = \sum_{nlj} (2j+1) v_{nlj}^2 \Psi_{nlj}^2(r)
\,.\label{eq:rhor}
\end{equation}
Using an interaction of zero range like (\ref{eq:vdelta}) leads to a definition
of a local gap-function
\begin{equation}
\Delta (r) = V(r) \chi(r)\,,\label{eq:localgap}
\end{equation} 
from which one can calculate the state-dependent pairing gaps
\begin{equation}
\Delta_{nlj} = \int \Delta(r) \Psi_{nlj}^2(r)\,r^2\,dr\,.
\label{eq:stategap}
\end{equation}
The BCS equations (\ref{eq:bcsm}) - (\ref{eq:stategap}) have to be solved
together with the HF equations in a self-consistent way. The Fermi energy
$\varepsilon_F$ is adjusted to reproduce the requested particle numbers or
densities for protons and neutrons. These equations can be reduced to an
approach, which we will call BCS with plane waves in the spherical box, by
restricting the radial wave functions for the single-particle states to the
spherical Bessel functions defined in (\ref{eq:basis1}) or (\ref{eq:basis2}).

\begin{figure}
\begin{center}
\epsfig{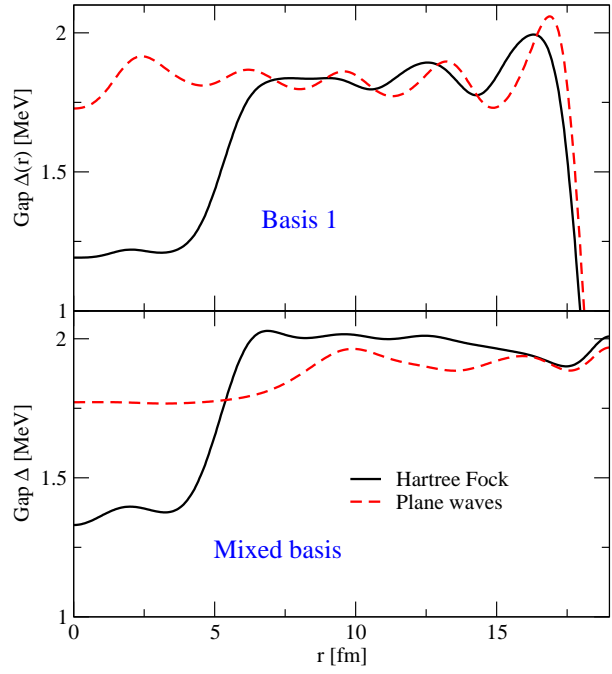}
\end{center}
\caption{\label{fig:deltar}Local gap function $\Delta (r)$ as defined in
(\protect{\ref{eq:localgap}}) for $\beta$-stable matter of density $\rho=0.02$
fm$^{-3}$. Results are displayed for the plane wave plus BCS approach (dashed
line) and the HF plus BCS approach (solid line) While the upper panel exhibits
results obtained in Basis 1 (see eq.(\protect{\ref{eq:basis1}}), 
the lower part of the figure displays results
obtained for the mixed basis.}
\end{figure}

Results for the gap function $\Delta (r)$ defined in (\ref{eq:localgap}) for
$\beta$-stable matter of a density $\rho$ = 0.02 fm$^{-3}$ are displayed in
Fig.~\ref{fig:deltar}. The gap function evaluated for the plane wave basis are
presented by the dashed line. We find that this gap-function fluctuates around
1.8 MeV with a sharp drop at the boundary of the spherical box. This drop is of
course related to the boundary condition (\ref{eq:basis1}) of all wave
functions. As already discussed for the local density above, this deficiency can
be cured by using a mixed basis, switching between (\ref{eq:basis1}) and
(\ref{eq:basis2}) for states with even and odd parity (see lower panel of
Fig.~\ref{fig:deltar}).

Solving the BCS plus Hartree-Fock equations one obtains a gap-function with
values which are suppressed by about 25 percent in the area of the quasi-nuclei.
From this result one may speculate that the inhomogeneous matter leads to regions
of enhanced densities in which the formation of pairing correlations is
suppressed to some extent. Therefore these regions might be considered as nuclei
for the formation of normal matter within the superfluid phase of neutron
matter. Also one could imagine that these quasi-nuclei could lead to vortex
pinning in the rotation of the superfluid crust of neutron stars.

\begin{figure}
\begin{center}
\epsfig{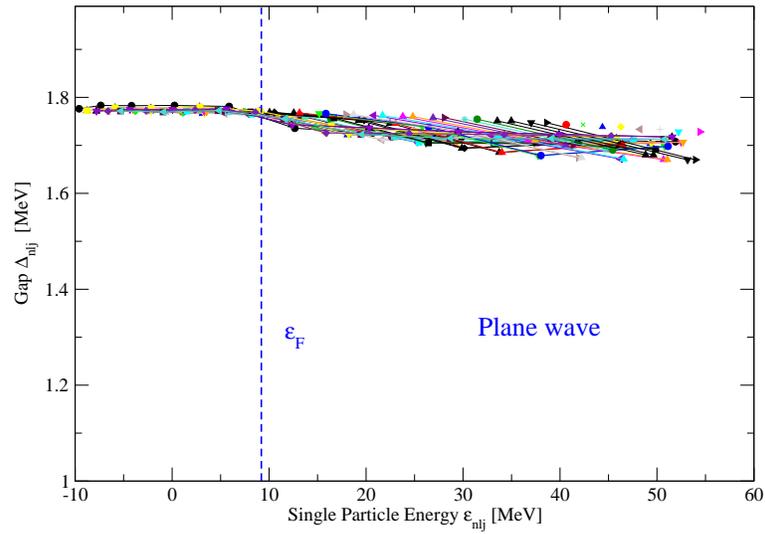}
\end{center}
\caption{\label{fig:deltapw}Results for the state-dependent pairing gaps
$\Delta_{nlj}$ defined in (\protect{\ref{eq:stategap}}) for $\beta$-stable 
matter of density $\rho=0.02$ fm$^{-3}$ as derived from plane wave plus BCS
calculation. These gaps are displayed as a function
of the corresponding $\varepsilon_{nlj}$. Each gap is presented by a symbol,
gaps belonging to states with identical orbital angular momentum $l$ and
angular momentum $j$ are connected by a line.}
\end{figure}

\begin{figure}
\begin{center}
\epsfig{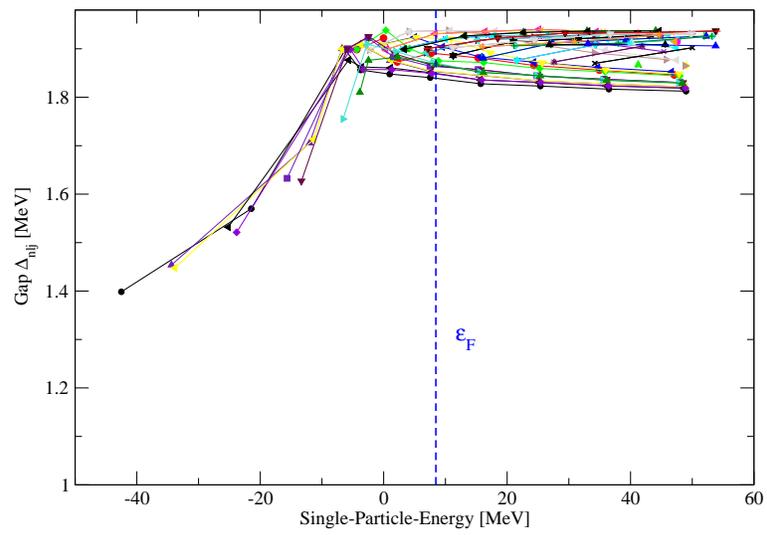}
\end{center}
\caption{\label{fig:deltahf}Results for the state-dependent pairing gaps
$\Delta_{nlj}$ as derived from a HF plus BCS
calculation. Further details see Fig.~\protect{\ref{fig:deltapw}}.}
\end{figure}

Before drawing any conclusions of this kind, we would like to compare the
results for the state-dependent gaps $\Delta_{nlj}$ displayed in
Figs.~\ref{fig:deltapw} and \ref{fig:deltahf} for the plane wave plus BCS and HF
plus BCS calculations, respectively. Comparing these results, one finds that
plane wave plus BCS calculation leads to a gap $\Delta_{nlj}$ which is almost
independent of the state. 

The calculation of the inhomogeneous system on the other hand leads to a few
single-particle states, which are much more bound than the lowest states in the
homogeneous calculation (-42 MeV as compared to -9.5 MeV). These deeply bound
states are localized near the origin. This large binding of the single-particle
states which leads to a reduction of the density of states is also responsible
for the reduction of the pairing gap for these low-lying states. 
On the other hand, the states with single-particle energies close to the Fermi 
energy $\varepsilon_F$ show a value for the gap parameter which is very close to
the one derived for the homogenous system. 

This implies that the reduction of
the gap-function $\Delta (r)$ in the region of the quasi-nuclei discussed above
manifests itself mainly in a reduction of the pairing gap for the deeply bound
single-particle states. The evaluation of response functions and other
observables, however, is in general sensitive to the gap at the Fermi energy.
Therefore one cannot expect major differences in the behavior of the
inhomogeneous as compared to the homogenous system of matter with respect to the
pairing properties of the neutrons. This shall be different for smaller
densities, where the Fermi energy for the neutrons drops to values at which the
shell structure of the single-particle energies is significant. 

\section{Conclusions}
The transition from homogenous to inhomogeneous matter as it occurs in the crust
of neutron stars has been investigated. Special attention has been paid to the
consequences of shell effects, which occurring assuming a phase of quasi-nuclei 
embedded in a sea of neutrons. For that purpose Hartree-Fock plus BCS
calculations are performed assuming a basis of single-particle states for a
spherical Wigner-Seitz cell. It is observed that the shell effects lead to a
substantial increase for the proton abundance in $\beta$-stable baryonic matter.

Shell effects are also responsible for a decrease of the localized pairing gap
in the region of the quasi-nuclei. A more detailed analysis, however, shows that
this reduction of the local gap function $\Delta(r)$ is mainly due to a
reduction of the pairing gap for the deeply bound single-particle states. The
pairing properties for the single-particle states close to the Fermi energy are
similar to those obtained for a homogenous description of the system.

The present studies are based on simple parameterizations for the effective NN
interaction. It may be of interest to study whether the predictions for the
transition from homogenous to inhomogeneous matter also hold, when more
realistic NN interactions are employed. 

The use of a spherical Wigner-Seitz cell is a source of various deficiencies.
The boundary conditions at the border of the cell lead to fluctuations in the
density, which complicates the comparison with the infinite homogenous system. 
This could be improved by employing a Cartesian basis, which is more involved
from the numerical point of view. The use of a Cartesian basis, however, would
also allow the study of non-spherical structures.

\bigskip
%\section{Acknowledgments}
We would like to thank Stephane Goriely, Mike Pearson and Mathieu Samyn
for very helpful discussions. One of us (F.M.) has been supported by a stipend of
the ``Deutsche Akademische Austauschdienst'' (DAAD, Germany). Also this
financial support is gratefully acknowledged

\vfil\eject

\begin{figure}
\begin{center}
\epsfig{figure=plot25.eps,width=10cm}
\end{center}
\caption{\label{plot25}The local density for homogeneous matter ($\rho=0.06$
fm$^{-3}$) calculated according to (\protect{\ref{eq:locdens}}) using Basis 1
(\protect{\ref{eq:basis1}}), Basis 2 (\protect{\ref{eq:basis2}}) and the mixed
basis as described in the text}
\end{figure}

\end{document}